\documentstyle[11pt,aaspp4]{article}
\tighten
\input psfig.tex

\begin{document}

\title{OSSE Observations of the Soft Gamma Ray Continuum from 
the Galactic Plane at Longitude $95^\circ$} 

\author{J. G. Skibo, W. N. Johnson, J. D. Kurfess, R. L. Kinzer, G. Jung and
J. E. Grove}

\affil{E. O. Hulburt Center for Space Research, Code 7653, Naval Research 
    Laboratory, Washington, DC 20375-5352}
 
\author{W. R. Purcell and M. P. Ulmer}

\affil{Northwestern University, Evanston, IL}

\author{N. Gehrels and J. Tueller}

\affil{NASA Goddard Spaceflight Center, Greenbelt, MD}

   
\begin{abstract}

We present the results of OSSE observations of the soft gamma ray continuum
emission from the Galactic plane at longitude 95$^\circ$. Emission is detected
between 50 and 600 keV where the spectrum is fit well by a power law with
photon index $-2.6\pm0.3$ and flux $(4.0\pm0.5)\times 10^{-2}$ photons s$^{-1}$
cm$^{-2}$ rad$^{-1}$ MeV$^{-1}$ at 100 keV.  This spectral shape in this range
is similar to that found for the continuum emission from the inner Galaxy but
the amplitude is lower by a factor of four. This emission is either due to
unresolved and previously unknown point sources or it is of diffuse origin, or
a combination of the two. Simultaneous observations with OSSE and smaller field
of view instruments operating in the soft gamma ray energy band, such as XTE or
SAX, would help resolve this issue. If it is primarily diffuse emission due to
nonthermal electron bremsstrahlung, as is the $>1$ MeV Galactic ridge
continuum, then the power in low energy cosmic ray electrons exceeds that of
the nuclear component of the cosmic rays by an order of magnitude. This would
have profound implications for the origin of cosmic rays and the energetics of
the interstellar medium. Alternatively, if the emission is diffuse and thermal,
then there must be a component of the interstellar medium at temperatures
$\sim10^9$ K.

\end{abstract}

\keywords{Gamma Rays: Observations --- ISM: General --- ISM: Cosmic Rays --- 
ISM: Supernova Remnants}

\eject

\section{Introduction}

The Galactic plane is an extended source of gamma radiation. This has been
demonstrated at energies above about 70 MeV with SAS-2 (Fichtel et al. 1975;
Hartman et al. 1979), COS-B (Mayer-Hasselwander et al. 1982; Strong et al.
1988) and most recently with EGRET on the Compton Gamma Ray Observatory (CGRO,
Hunter et al. 1995; 1997). This high energy emission is generally believed to
result from cosmic-ray interactions with interstellar gas and radiation through
the processes of electron bremsstrahlung, inverse Compton scattering and pion
production (e.g. Bloemen 1989; Bertsch et al. 1993). Observations made with
COMPTEL on CGRO (Bloemen et al. 1993; Strong et al. 1994; 1995; 1996) have
confirmed that this Galactic plane emission extends down to energies near 1
MeV.  Whereas there is no proof that this emission is genuinely diffuse and not
the superposition of unresolved point sources, the spectrum of this emission
from the Galactic center direction is a smooth extrapolation down to 1 MeV of
the continuum measured with SAS-2, COS-B and EGRET which are fit adequately
with diffuse emission models. 

The diffuse Galactic soft gamma ray continuum is currently not well determined
below 1 MeV.  This is because at these energies the Galactic continuum is
complicated by the presence of a number of point sources, many of which are
variable. This energy range is, however, a very interesting energy regime for
continuum processes. For, below 1 MeV, the candidate radiation mechanisms
[inverse compton scattering, nonthermal bremsstrahlung from cosmic ray
electrons or thermal bremsstrahlung from galactic ridge thin hot ($T\sim10^8$
K) plasma] could, in principle, contribute in approximately equal amounts to
the total emission. 

Observations of the Galactic plane toward the Galactic center made with the
OSSE instrument on CGRO display evidence for daily variability at energies
$<150$ keV (Ulmer et al. 1997). The inferred variability implies that a
significant portion of the Galactic plane continuum measured with OSSE from
this direction is from point source emission. In another analysis of Galactic
plane observations made with OSSE (Kurfess 1995, Purcell et al. 1996) it was
found that, when the contribution from the prominent point sources monitored
during simultaneous observations with SIGMA is subtracted from the Galactic
center spectrum measured with OSSE, the diffuse emission is essentially
identical to that measured from the Galactic plane at $\ell=25^\circ$.
Furthermore, the residual source-subtracted spectrum of this emission changes
from a photon index of $\Gamma\simeq2.7$ ($dN/dE\propto E^{-\Gamma}$) for
energies below $\sim 200$ keV to $\Gamma\simeq1.7$ at higher energies (Purcell
et al. 1996; Strong et al. 1994). Thus, the soft gamma ray emission (below 200
keV) from the inner Galaxy is more intense than the extrapolation of the higher
energy continuum with index $\Gamma\simeq1.7$ and has a roughly constant
longitude distribution over the central radian of the Galaxy when the prominent
sources are removed. 

This corroborates observations made with the low-energy gamma-ray experiment
flown on HEAO-1 (Peterson et al. 1990), where it was shown that the 90 - 280
keV continuum emission has a longitude distribution similar to that
$\gtrsim100$ MeV. More recently, Gehrels et al. (1991) have shown that the low
energy (30-200 keV) spectrum of the Galactic center observed with the balloon
borne Ge spectrometer GRIS was at that time very nearly equal to the spectrum
observed from $\ell=335^\circ$, after correcting the Galactic center spectrum
for the contributions from the point sources 1E1740.7-2942 and GRS1758-258.
Furthermore, hard X-ray observations from the direction of the Galactic Center
with detectors of moderate fields of view (15$^\circ$ to 30$^\circ$) show that,
even though the observed fluxes vary in time, they have a lower envelope which
coincides with the observed flux from $\ell=335^\circ$ (Gehrels \& Tueller
1993). In addition, emission from the Galactic plane has been detected through
gaps in the passive collimators of the SIGMA telescope with
essentially the same spectrum as observed with the other instruments (Claret et
al. 1995). 

At X-ray energies extended emission has been observed from the Galactic ridge
with {\it Ginga} (Koyama et al. 1989; Yamasaki et al. 1996). The spectrum
displays prominant 6.7 keV line emission from He-like Fe and a continuum that
is fit by a thermal bremsstrahlung with a hint of a high energy tail around 10
keV. This emission is probably a mixture of thermal and nonthermal
bremsstrahlung. Observations of the diffuse X-ray emission (2.5-22 keV) with
ART-P onboard {\it Granat} from a region a few degrees around the Galactic
center show that the hard emission (8.5-22 keV) has a different morphology,
being more extended along the Galactic disk than the soft X-ray emission
(Markevitch, Sunyaev, \& Pavlinsky 1993; Sunyaev, Markevitch, Pavlinsky 1993). 

Information on the gamma ray spectrum from regions outside the central radian
of the Galaxy is sparse because the overall intensity is much lower. In this
paper we present the results of an analysis of OSSE observations of the
Galactic plane at $\ell=95^\circ$. This is a direction nearly tangent to the
solar circle where there are few catalogued hard X-ray sources. We find that
the continuum below 200 keV has the same general spectral characteristics as
observed by OSSE from the inner Galaxy, although it is reduced in intensity by
a factor of 4. The spectrum below $\sim 600$ keV is adequately fit by a power
law with photon index $-2.6\pm0.3$. At this longitude OSSE does not have
sufficient sensitivity to detect the expected harder high energy component. If
this emission is truly of diffuse origin and produced by cosmic rays
interacting with ambient gas and radiation, then the spectrum and Galactic
radial extent measured by OSSE impose severe energetic constraints on cosmic
ray origin scenarios. 
  
\section{Observations and Analysis}

The Oriented Scintillation Spectrometer Experiment (OSSE) consists of 4 nearly
identical Na(Tl)/CsI(Na) phoswich detectors operating in the 50 keV - 10 MeV
energy range. Each detector is passively collimated with a tungsten slat
collimator defining an optical aperature (field of view) of
$\sim11.4^\circ\times3.8^\circ$ (FWHM) (see Johnson et al. 1993 for details). 

Observations of the Galactic plane at longitude 95$^\circ$ were carried out in
1996 during viewing periods 515 (February 20 - March 5) and 519 (April 23 - May
7). The detector orientations at these times is shown in Figure 1. Each
detector alternately observed the plane and background fields located
$\pm9^\circ$ (period 515) and $\pm12^\circ$ (period 519) on alternating sides
of the target field along the direction perpendicular to the long axis of the
collimator (see Figure 1). During period 515 the detector orientation was such
that the long axis of the collimator was nearly aligned with the galactic
plane, while in period 519 it was inclined at $40^\circ$. The background
corrected spectra represent the excess of the two target fields over the
corresponding background fields. Small corrections to these difference spectra
for systematic scan-angle dependent background effects were made.

The background subtracted count rate at longitude $95^\circ$ (period 515) was
about a factor of four lower than that obtained from OSSE measurements of the
inner Galaxy. This can be compared with observations made at higher energies
($\sim100$ MeV) with COS-B where the flux level drops off by only a factor of
about three in going from the Galactic center to the Solar tangent. 

The background subtracted count rate summed over all 4 detectors in 1 day time
intervals from each of the viewing periods separately showed no evidence for
variability within either observation period. The 50-150 keV count rates were
$3.2\pm0.3$ counts s$^{-1}$ MeV$^{-1}$ for period 515 and $1.7\pm0.3$ counts
s$^{-1}$ MeV$^{-1}$ for period 519. This is a difference of about $3.5\sigma$
and is consistent with that expected from the reduced exposure to the plane
during period 519 and the fact that the background fields for period 519
partially overlap the plane (see Fig. 1). For a line-like source in the plane
with negligible latitude extent ($\lesssim1^\circ$) the measured flux in period
519 would be approximately 50\% that of period 515. This fraction increases to
about 70\% if a wider latitude distribution ($\approx5^\circ$ Gaussian FWHM) is
assumed. From the measured count rates this fraction is determined to be
$0.54\pm0.11$. This value places a a $2\sigma$ upper limit of 
FWHM$\approx8^\circ$.

To convert from flux through the OSSE collimator to flux per radian it was
necessary to assume a spatial distribution for the emission. We assumed that
the longitude distribution at longitudes ($85^\circ$-$105^\circ$) is roughly
constant over the OSSE field of view. This is as expected for diffuse emission
based on the $\gtrsim30$ MeV EGRET observations (Hunter et al. 1997).
Furthermore, we assumed that the distribution has the same latitude extent as
that measured with OSSE from the inner Galaxy where the distribution is fit
well by a gaussian with FWHM$\approx5^\circ$ (Purcell et al. 1996). We
acknowlege that the true latitude distribution cannot be determined from our
data which only provide the $2\sigma$ upper limit of $8^\circ$. If, for
example, FWHM$\lesssim1^\circ$ then the flux per radian would be lower by a
factor of 0.67. On the other hand, if FWHM$\approx8^\circ$ then the flux per
radian would be larger by a factor of 1.4. 

The spectra from each viewing period are adequately fit by single power law
models of spectral index $-2.7\pm0.3$ and $-2.3\pm0.6$ for periods 515 and 519,
respectively. Within the errors these are consistent. To combine the data from
both viewing periods we scale the count rate of period 519 by a factor which
renders the 50$-$150 keV count rates of both periods identical. This compensates
for the different collimator orientations. These corrections are valid only up
to 600 keV where the data become statistically uncertain. Above that energy,
where the current observations are marginally significant, the OSSE collimator
response must be modeled more carefully. 

We find that the spectrum over the range 50-600 keV in the combined data is
fit well by a single power law,
$\Phi(E)=A(E/0.1{\rm\,MeV})^{-\Gamma}$, where
$A=(4.0\pm0.5)\times10^{-2}$ photons s$^{-1}$ cm$^{-2}$ rad$^{-1}$ MeV$^{-1}$
and $\Gamma=2.6\pm0.3$. The reduced $\chi^2$ for 88 degrees of freedom is 1.15,
corresponding to a probability of 0.15. In Figure 2 we show this fit with the
data. 

Other spectral models, such as a broken power law or thermal Comptonization
plus power law, cannot be ruled out, but do not provide significantly improved
fits. Fitting to a single power law restricted to energies below 200 keV
results in a spectral index $\Gamma=2.7\pm0.3$. The fit obtained using a broken
power law has a break near 200 keV but is identical within errors below the
break to that of the single power law shown in Figure 2. The spectral index
above the break is harder but not well constrained. It is in accord with the
spectrum measured with OSSE from the inner Galaxy (Kurfess 1995; Purcell et al.
1996). 

An acceptable fit is also obtained with an optically thick thermal
Comptonization (Sunyaev \& Titarchuk 1980) plus power law model. The
Sunyaev-Titarchuk component required to account for the $\lesssim200$ keV
emission has temperature in the range 10-30 keV and a Thomson depth that is not
very well constrained. The high energy power law is hard $\Gamma\lesssim-2$ but
not meaningful due to the limited statistical significance of the data above
200 keV. Such a model might represent the emission from an X-ray binary or
distribution of such sources with thermal Comptonization spectra superposed on
a background diffuse power law component. 

\section{Origins and Implications}

The central question concerning the origin of this soft gamma ray emission is
whether it is primarily diffuse or if it is greatly affected by one or more
compact sources. There are 2 candidate hard X-ray point sources, the eclipsing
LMXB 4U 2129+47 located at $l=91.58^\circ$, $b=-3.04^\circ$ (Thorstensen et al.
1979; Ulmer et al. 1980; Garcia 1994) and the X-ray pulsar Cep X-4 at
$l=99.08^\circ$, $b=3.78^\circ$ (Ulmer et al. 1973; Koyama et al. 1991). As can
be seen from Figure 1 the detectors were more sensitive to both of these
sources during viewing period 519. Hence, if either of these sources were
substantial contributors to the observed OSSE flux then the flux observed in
viewing period 519 would have exceeded that observed in 515 in the absence of
source variability. However, the 50-150 keV count rate in viewing period 519
was found to be 54\%$\pm$11\% that of 515, more in accord with the emission 
being distributed along the Galactic plane. 

In the range 50-150 keV the OSSE collimator is fairly well represented by a
$3.8^\circ\times11.4^\circ$ FWHM triangular response function. In viewing
period 515, only 4U 2129+47 was in the detector field of view and only at the
7\% response level, hence we rule out Cep X-4 as a potential contributor to the
viewing period 515 flux. In viewing period 519, 4U 2129+47 was in the field of
view at the 54\% level. We estimate the maximum contribution of 4U 2129+47 to
the viewing period 515 flux in the 50-150 keV band. Assuming a latitude extent
of $5^\circ$ for the diffuse emission, the ratio of the response to the plane
in viewing period 519 to that of 515 is 69\%. From these numbers a $2\sigma$
upper limit of 2\% can be placed on fraction of the 50-150 keV emission
contributed by 4U 2129+47 in viewing period 515. Even if we make the unlikely
assumption that the diffuse emission has neglegible latitude extent
($\lesssim1^\circ$), then the ratio of the response to the plane in viewing
period 519 to that of 515 is 52\%, which allows for a $2\sigma$ upper limit of
3\% for the contribution of 4U 2129+47 in viewing period 515. 

Of course, suitably arranged time variability can be contrived to yield an even
greater contribution from 4U 2129+47. However, because each observation
separately showed no evidence for day to day variability, this scenario seems
unlikely. In addition, if 4U 2129+47 was responsible for the viewing period 515
emission then its intrinsic flux corrected for the collimator response would be
approximately 100 mCrab, an order of magnitude higher than the observed OSSE
flux in period 515. In this case it would have been detected with BATSE on
CGRO. In summary, although the ratio of the 50-150 keV count rate measured in
viewing period 515 to that in 519 is consistent with the emission coming solely
from the plane, we cannot rule out that 3\% ($2\sigma$ upper limit) is due to
4U 2129+47. Furthermore, we cannot rule out that the observed fluxes are due to
other previously undetected hard X-ray sources. For the observed fluxes to be
produced by a single source of constant intensity, it is constrained to lie in
the darkly shaded region of Figure 1 and be at least 10 mCrab intensity in the
50-150 band. A source at this low intensity would not have been detected by 
BATSE over the duration of the OSSE observation.

Another possibility is that this emission arises from supernova remnants
(SNR's). Recent X-ray observations with ASCA of the supernova remnant SN 1006
reveal intense nonthermal X-ray emission arising from the edges of the remnant
shell (Koyama et al. 1995). This has been interpreted as synchrotron emission
from $\gtrsim100$ TeV electrons accelerated by the first order Fermi mechanism
in the blast wave shock (Reynolds 1996). In fact, a nonthermal tail has been
observed with OSSE from the supernova remnant Cas A (The et al. 1995). There
were two SNR's in the fields of view of the Galactic plane $l=95^\circ$
pointing, CTB 104A and 3C 434.1, neither of which has been detected at X-ray
energies (Green 1995). 

If, on the other hand, this emission is genuinely of a diffuse origin, then it
could either be bremsstrahlung from low energy ($\lesssim$ 1 MeV) cosmic ray
electrons or inverse Compton emission from high energy electrons. To Compton
up-scatter photons from the cosmic microwave background, ambient infrared and
starlight photons to energies $\sim100$ keV requires electrons of energy in the
range 100 MeV to 10 GeV. Electrons at these energies are responsible for
producing the Galactic synchrotron radio emission at frequencies $\sim100$ MHz
(e.g. Berezinskii et al. 1990). However, the radio spectrum is observed to
steepen above $\sim100$ MHz. This is not in accord with OSSE observations of
the inner Galaxy which display the break in the opposite sense. 

For these reasons we next consider bremsstrahlung as a more likely mechanism.
There is a problem, however, with the bremsstrahlung interpretation.
Bremsstrahlung of $\lesssim 1$ MeV electrons is energetically highly
inefficient. The energy loss rate due to ionization and Coulomb collisions with
the ambient plasma is several thousand times greater than the loss rate due to
bremsstrahlung. On the Galactic scale a large power is required to maintain
these sub-MeV cosmic ray electrons against energy losses to the interstellar
medium. Since these electrons do not escape the Galactic disk, the Galaxy acts
as a sort of calorimeter (Pohl 1993). The Galactic soft gamma ray continuum
luminosity is estimated to be $\sim 10^{38}$ erg s$^{-1}$ (Skibo, Ramaty, \&
Purcell 1995; 1996) from OSSE measurements of the inner Galaxy (Purcell et al.
1996). Therefore, more than $10^{41}$ erg s$^{-1}$ must be supplied to the
electrons producing this radiation (Chi \& Wolfendale 1991; Skibo \& Ramaty
1993; Skibo, Ramaty, \& Purcell 1995; 1996). This exceeds by approximately an
order of magnitude the power supplied to the nuclear component of the cosmic
rays (e.g. Berezinskii et al. 1990). Furthermore, it is approaching the total
power supplied by supernovae, the traditional sources of the Galactic
cosmic-rays ($\sim10^{42}$ erg s$^{-1}$ assuming $10^{51}$ erg every 30 years).
This large power is efficiently deposited into the ISM in the form of heat and
ionization, thus these electrons could be an important component. For example,
they could provide some of the required $10^{41}$ erg s$^{-1}$ in heat and
ionization to maintain the warm ($10^4$ K) intercloud medium (Reynolds 1990;
Skibo, Ramaty, \& Purcell 1995). However, if these electrons penetrate cold
neutral clouds then excessive molecular dissociation would occur (Skibo,
Ramaty, \& Purcell 1995). 

It has recently been shown (Schlickeiser 1996) that stochastic Fermi
re-acceleration by gyro-resonant wave-particle interactions in the ISM can give
rise to the steep electron spectrum as required by the gamma ray observations.
However, the total power of $\gtrsim10^{41}$ erg s$^{-1}$ must still be
supplied to the electrons through hydromagnetic turbulance. Whether such power
is availiable in this form is an open question. 

If, on the other hand, the emission is thermal bremsstrahlung from hot thin
interstellar plasma then the power required is simply that which is radiated,
$\sim10^{38}$ erg s$^{-1}$. This is far less demanding energetically, but it
implies temperatures $\sim 10^9$ K. It is not clear whether gas in the ISM can 
be maintained at these temperatures, but detailed modeling of the ISM with 
such a super hot component is beyond the scope of this work.

\section{Summary}

Gamma ray continuum emission has been detected between 50 and 600 keV with the
OSSE instrument on CGRO from the Galactic plane at longitude $95^\circ$. The
spectrum is fit well by a power law with photon index $-2.6\pm0.3$ and flux
$(4.0\pm0.5)\times 10^{-2}$ photon s$^{-1}$ cm$^{-2}$ rad$^{-1}$ MeV$^{-1}$ at
100 keV.  This spectrum is consistent with that of the soft gamma ray component
of the diffuse continuum emission from the inner Galaxy but is lower by
approximately a factor of four in intensity. We conclude that either this
emission is due to unresolved and previously unknown point sources or that it
is diffuse electron bremsstrahlung, or a combination of the two. To resolve
this issue simultaneous observations with OSSE and smaller field of view
instruments operating in the soft gamma ray energy band, such as the X-ray
Timing Explorer (XTE) or Beppo-SAX. If the emission is nonthermal
bremsstrahlung then a very large power is required, whereas a thermal
bremsstrahlung interpretation requires temperatures $\sim10^9$ K. 

\acknowledgments 

We thank Kellie McNaron-Brown for help with the OSSE data analysis. 
This work was supported through NASA grant S-67001-F.  
This research has made use of data obtained through the High Energy
Astrophysics Science Archive Research Center Online Service, provided by
the NASA/Goddard Space Flight Center.

\clearpage

\begin{figure*}
\centerline{\psfig{figure=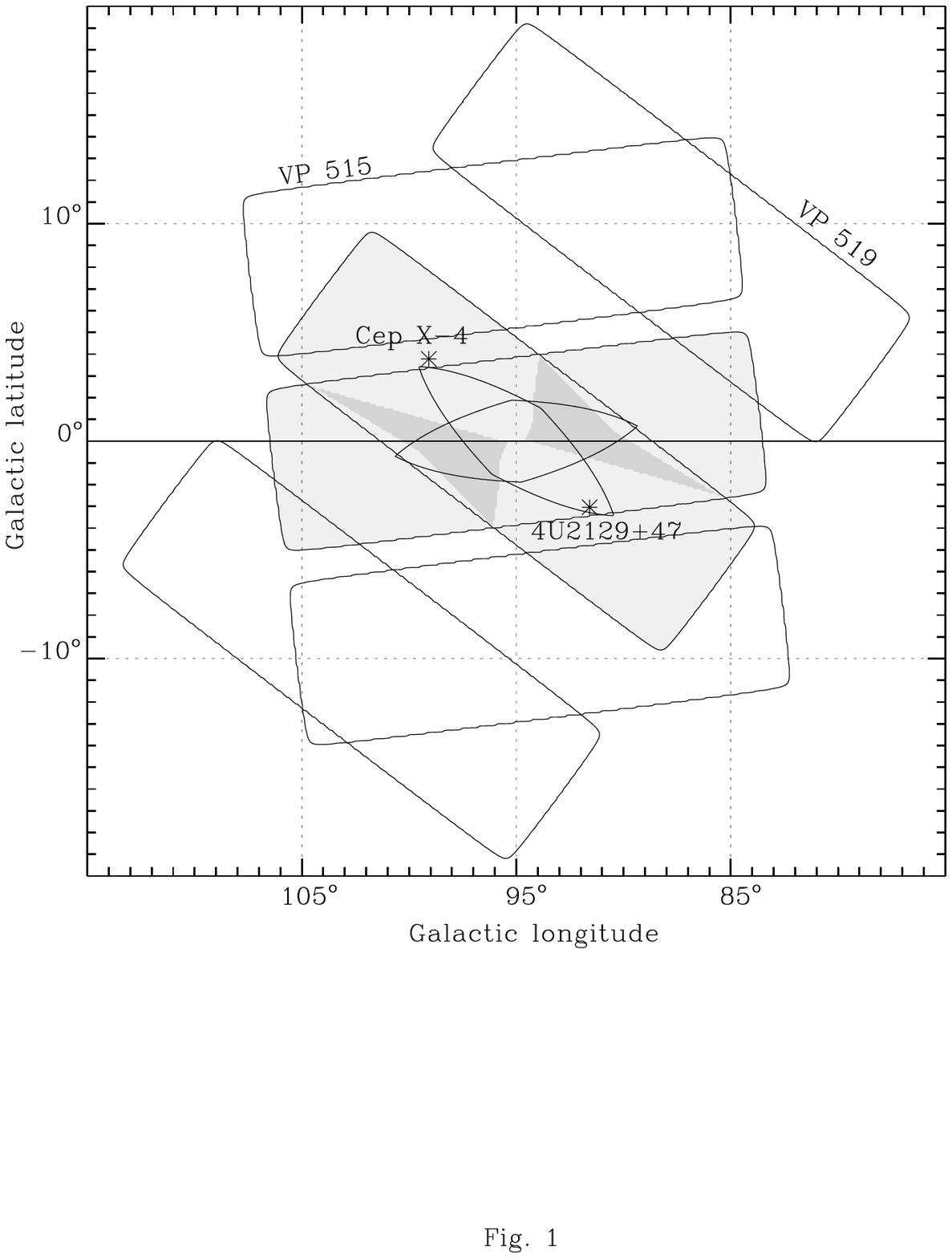,height=10 cm,width=15 cm}}
\vskip 0.8truecm
\caption{ \em
Scematic of the OSSE viewing strategy of the Galactic plane at longitude
$95^\circ$ during viewing periods 515 and 519. The rectangles represent the
full OSSE field of view ($11.4^\circ\times3.8^\circ$). Each observation
consists of three boxes: the target (shaded) and two background fields on
either side (unshaded). The diamond shaped contours in the target fields
represent the 50\% collimator response level. The dark shaded region is where a
single point source must lie if it is responsible for all of the emission. 
}
\end{figure*}

\begin{figure*}
\centerline{\psfig{figure=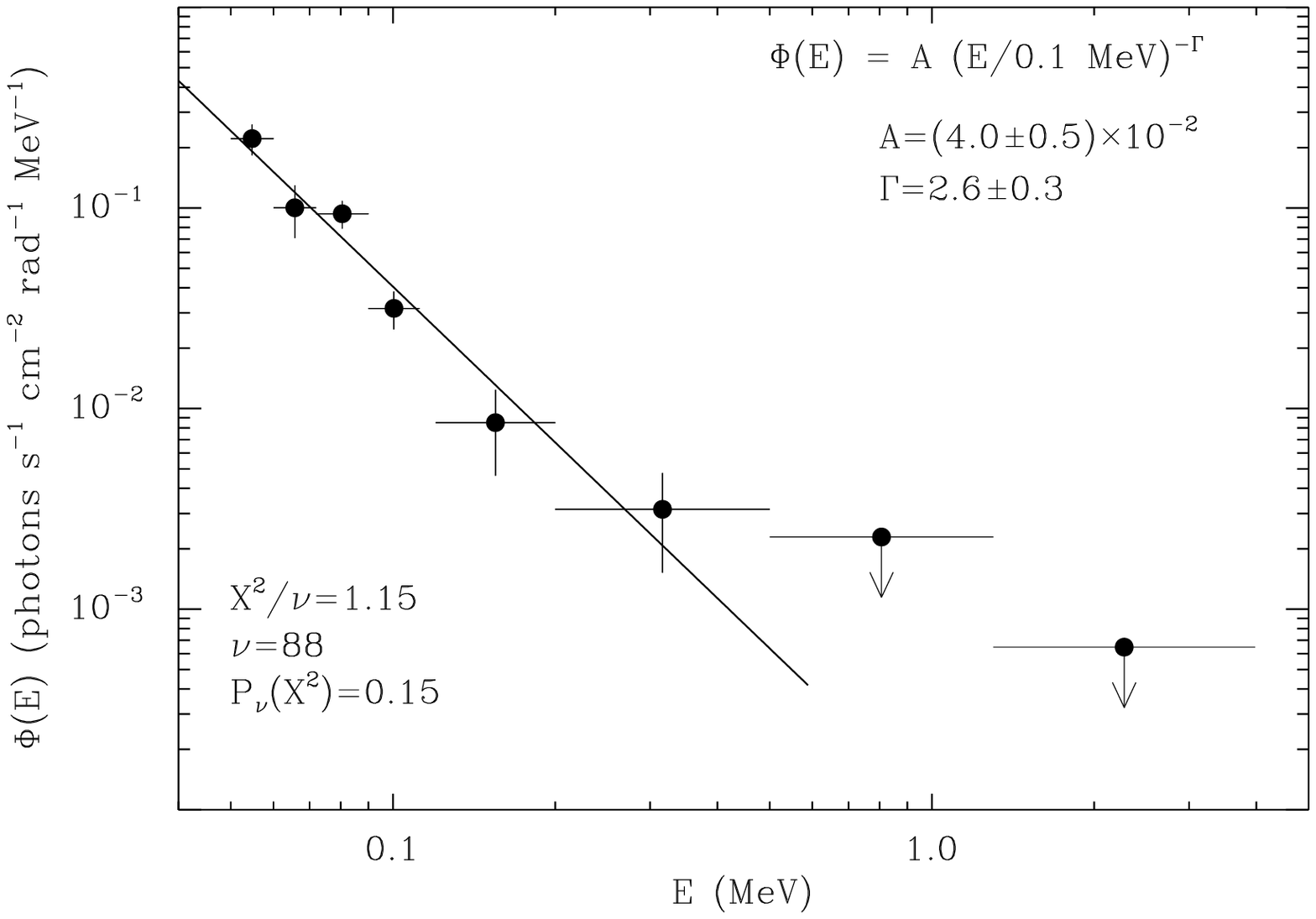,height=10 cm,width=15 cm}}
\caption{ \em
OSSE spectrum from the direction $\ell=95^\circ$ (viewing periods 515 and 519
combined) and single power law fit. 
} 
\end{figure*}


\begin{references}

\reference{} Berezinskii, V. S. et al. 1990, Astrophysics of Cosmic
Rays (North Holland, Amsterdam), 191

\reference{} Bertsch, D. L. et al. 1993, \apj, 416, 587

\reference{} Bloemen, J. B. G. M. 1989, ARA\&A, 27, 469

\reference{} Bloemen, J. B. G. M. et al. 1993, Compton Observatory Science
Symposium, eds. M. Friedlander, N. Gehrels \& D. Macomb (New York:
AIP), 30

\reference{} Chi, X. \& Wolfendale, A. W. 1991, in The Interstellar Disk-Halo
Connection in Galaxies, ed. H. Bloemen (IAU Symp. 144), 197

\reference{} Claret, A. et al. 1995, Adv. Space Res., 15, (5)57

\reference{} Fichtel, C. E. et al. 1975, \apj, 198, 163

\reference{} Garcia, M. R. 1994, \apj, 435, 407

\reference{} Gehrels, N. et al. 1991, \apj,  375, L13

\reference{} Gehrels, N. \& Tueller, J. 1993, \apj, 407, 597

\reference{} Green D.A., 1995, A Catalogue of Galactic Supernova Remnants
(1995 July version), Mullard Radio Astronomy Observatory,
Cambridge, United Kingdom

\reference{} Hartman, R., Kniffen, D., Thompson, D., \& Fichtel, C. 1979, ApJ,
230, 597 

\reference{} Hunter, S. D. et al. 1995, 24th Internat. Cosmic Ray Conf., 182

\reference{} Hunter, S. D. et al. 1997, \apj, in press

\reference{} Johnson, W. N. et al. 1993, \apjs, 86, 693

\reference{} Koyama, K. et al. 1989, Nature, 339, 603

\reference{} Koyama, K. et al. 1991, \apj, 366, L19

\reference{} Koyama, K. et al. 1995, Nature, 378, 255

\reference{} Kurfess, J. D. 1995, 17th Texas Symposium on Relativistic
Astrophysics (New York Academy of Sciences)

\reference{} Levine, A. M. et al. 1984, \apjs, 54, 581

\reference{} Markevitch, M., Sunyaev, R. A., \& Pavlinsky, M. 1993, Nature, 
364, 40

\reference{} Mayer-Hasselwander, H. A. et al.  1982, A\&A, 105, 164

\reference{} Peterson, L. E. et al. 1990, 21st Internat. Cosmic Ray Conf., 
1, 44

\reference{} Pohl, M 1993, A\&A, 270, 91

\reference{} Purcell, W. R. et al., 1995, 24th Internat. Cosmic Ray Conf., 
211

\reference{} Purcell, W. R. et al. 1996, A\&AS, 120, 389

\reference{} Reynolds, R. J. 1990, ApJ, 349, L17

\reference{} Reynolds, S. P. 1996, \apj, 459, L13

\reference{} Schlickeiser, R. 1996, The Physics of Galactic Halos, eds. 
H. Lesch, R. J. Dettwar. U. Mebold, \& R. Schlickeiser (Akademie-Verlog,
Berlin)

\reference{} Skibo, J. G. \& Ramaty, R. 1993, A\&AS, 97, 145

\reference{} Skibo, J. G., Ramaty, R., \& Purcell, W. R. 1995, 24th Internat. 
Cosmic Ray Conf., 219

\reference{} Skibo, J. G., Ramaty, R., \& Purcell, W. R. 1996, A\&AS, 120, 403

\reference{} Strong, A. W. et al. 1988, A\&A, 207, 1

\reference{} Strong, A. W. et al. 1994, A\&A, 292, 82

\reference{} Strong, A. W. et al. 1995, 24th Internat. Cosmic Ray Conf., 234

\reference{} Strong, A. W. et al. 1996, A\&AS, 120, 381

\reference{} Sunyaev, Markevitch, M., \& Pavlinsky, M. 1993, \apj, 407, 606

\reference{} Sunyaev, R. A. \& Titarchuk, L. G. 1980, \aap, 86, 121

\reference{} The, L. S. et al. 1995, \apj, 444, 244

\reference{} Thorstensen, J. et al. 1979, \apj, 233, L57

\reference{} Ulmer, M. P. et al. 1973, \apj, 184, L117

\reference{} Ulmer, M. P. et al. 1980, \apj, 235, L159

\reference{} Ulmer, M. P. et al. 1997, The Transparent Universe, in press 

\reference{} Yamasaki N., et al., 1996, A\&AS, 120, 393

\end{references}
\end{document}